\documentclass[journal,9pt]{IEEEtran}
\usepackage{amsmath,amsfonts}
\usepackage{algorithmic}
\usepackage{algorithm}
\usepackage{array}
\usepackage{xcolor}

\usepackage[caption=false,font=normalsize,labelfont=sf,textfont=sf]{subfig}
\usepackage{textcomp}
\usepackage{stfloats}
\usepackage{url}
\usepackage{verbatim}
\usepackage{graphicx}
\usepackage{cite}
\usepackage{hyperref}
\newcommand{\Em}{{{\bf E}}}
\newcommand{\Zbf}{\mathbf{Z}}

\newcommand{\Eoc}{\Em^{{\rm oc}}}

\newcommand{\Ezl}[1]{\Em^{\mathbf{Z}_{{\rm L}#1}}}
\newcommand{\Ezll}[1]{\Em^{\mathbf{Z}_{\rm L}^{#1}}}

\begin{document}

\title{\huge{Non-uniform Antenna Loading Effect on Embedded Element Patterns and Application to Fault Detection}}

\author{\Large{Georgios Kyriakou}
        % <-this % stops a space
\thanks{Received 25 January 2026; revised 10 June 2026; accepted 8 July 2026.
The author is with the Department of Physics, University of Rome ‘La
Sapienza’, 00185 Rome, Italy (e-mail: georgios.kyriakou@uniroma1.it).
This article has supplementary downloadable material available at
https://doi.org/10.1109/TAP.2026.3713390, provided by the authors.
Digital Object Identifier 10.1109/TAP.2026.3713390}}% <-this % stops a space
%\thanks{Manuscript received Month X, 2026; revised Month XX, 202X.}

% The paper headers
\markboth{IEEE Transactions on Antennas and Propagation,~Vol.~X, No. X, 202X}%
{Shell \MakeLowercase{\textit{et al.}}: A Sample Article Using IEEEtran.cls for IEEE Journals}

%\IEEEpubid{0000--0000~\copyright~2023 IEEE}
% Remember, if you use this you must call \IEEEpubidadjcol in the second
% column for its text to clear the IEEEpubid mark.

\maketitle

\begin{abstract}
An iterative {technique} is presented to calculate the Embedded Element Pattern (EEP) transformation from a set of patterns computed for a uniform antenna port loading (scaled identity matrix) to a set of those computed for a non-uniform one (arbitrary diagonal matrix). {Inverting this transformation} to derive the non-uniform {impedances} of the arbitrary load {leads to a new, efficient algorithm which }disposes of the redundancy of {the trivial use of all \( N \) EEPs at the faulty array condition, requiring only one such EEP for numerically stable impedance fault calculation}. As the EEPs are envisioned to be obtained primarily through measurement, our method is also tested with the inclusion of various noise components and its convergence is evaluated, {leading to suggested measurement apparatus} SNR and fading levels, as well as the optimal choice of reference antenna to minimize the estimation error.
\end{abstract}


\section{Introduction}
\IEEEPARstart{C}{haracterisation} of antenna arrays for radio astronomy in terms of Embedded Element Patterns (EEPs) under different loading conditions and their subsequent application to beam steering is a well-developed theory \cite{warnick2021}. Inverse problems, such as finding the loading condition that leads to specific, usually measured, EEPs are also of interest. This is an important case for fault diagnosis, as the front-end on which a phased array is mounted can vary because of the failure of low-noise amplifiers (LNAs). This, in turn, is a consequence of environmental effects that are common in operation fields and lead to electrical and mechanical damage or other technical problems.

While some solutions have been presented in the literature to identify faulty antennas \cite{prajosh2022,rodriguez2009} using EEPs, a practical algorithm to the problem of calculating faulty termination impedances loading the array has not yet been devised. The EEP transformations from an ideal to an unknown loading condition of an array can be easily inverted to obtain the latter one {directly from \cite[Eq.~(1)]{kovaleva2020}}, but for antenna ports without cross-talk (disjoint antennas) only the \( N \) diagonal elements of an impedance load matrix need to be recovered. This practically means there is redundancy in the computation when calculating \( N \) non-zero elements using \( N\times M \) matrices, where \( M \) is the number of simulation or measurement points in the 3D far-field patterns. {In \cite{prajosh2022}, the authors do use a single measurement probe significantly reducing the input data required in their method, but only `on-off' faults are recovered by means of such a method. The same limitations apply to \cite{rodriguez2009} as well, where an approximation of mutual coupling employing single-fault EEPs is also considered further compromising accurate recovery.}

To tackle this {redundancy} issue, a perturbative linear algebra derivation has proven useful in computing the new EEPs under faulty termination. Recently, Buck \emph{et~al.} \cite{buck2023} showed that a more complicated relationship arises when each EEP has to be calculated with a different source impedance than that of the passive loading condition of all other elements. Such a practice is also applicable if one of the passive elements is terminated diversely than all the others, as well as than the source impedance of the EEP at hand. This was comprehensively shown in \cite{kyriakou2024}. 

In this communication, we revisit this formulation and generalize it employing the loaded array admittance matrix with an aim to connect the {\it nominal} EEPs (loaded with a uniform impedance at all array elements) to the {\it faulty} EEPs (loaded with an arbitrary impedance {at each array element)}. This is achieved by a recursive approach of rank-one updates of both the admittance transformation when a new fault becomes present and the new EEP calculation, when the array is loaded with the updated admittance matrix. {When calculations are inverted, we obtain an algorithm that estimates impedance faults bearing the novel aspect of requiring only one faulty EEP for accurate computations, a significant \( N\)-fold reduction in input data.} 

From a practical point of view, Unmanned Aerial Vehicle (UAV) measurements of EEPs have also become an appealing method of aperture array testing, the more such systems become cheaper and more accurate \cite{kandregula2024}. Aperture arrays for low-frequency radio astronomy, such as the Square Kilometer Array (SKA) and its precursors, could be diagnosed for front-end impedance mismatches using the proposed method, using existing UAV systems \cite{ciorba2024,weng2025} without the need for multiple antennas being measured, by using, instead, only a reference one. This constitutes an {in-situ} advantage, since reference elements are used for multiple purposes in such arrays, {even dealing with challenges such as accurate phase acquisition \cite{sanchez2020}}. In order for our method to be applicable in this case, robustness with respect to measurement noise has to be demonstrated, similarly as in \cite{kovaleva2023,kyriakou2024b}.

The communication is structured as follows: in Sec.~\ref{sec:eep_faulty}, we review the calculation of \cite{kyriakou2024}, its network equivalent and its inversion. We then generalise that calculation for \( N\) faults in Sec.~\ref{sec:new_algorithm}, where we introduce our algorithm. In Sec.~\ref{sec:simulation_nonoise} the algorithm is verified using the EEPs of a 16-element tile of the Murchison Widefield Array, while in Sec.~\ref{sec:simulation_noise}, we introduce additive noise and a fading channel between the measurement probe and the array, and repeat the algorithm calculations for various SNR values (additive noise) as well as channel fading levels (multiplicative noise) to establish its convergence rate and precision to correctly predicting the faulty terminations. Sec.~\ref{sec:conclusions} presents our conclusions.

\section{EEPs with Single Faulty Termination and Inverse Impedance Equation \label{sec:eep_faulty}}

We begin by stating our notation conventions: boldface is used for vectors and matrices (of each resulting object, after indexing). {Indexing of matrix columns is performed as a single variable subscript, and since we will only work with symmetric network matrices, indexing of rows will be denoted with the transposed column. Indexing of both the row and column is denoted by two variables in the subscript.} The iteration number of an object is denoted as a superscript. Finally, overbar electric field quantities are vector stacking at sampled far-field points. \( \text{diag}(\cdot) \) means the diagonal matrix of the argument, \( (\cdot)^{\rm T}\) is the matrix-transpose {and \((\cdot)^{\rm H}\) is the Hermitian of a matrix}.

\begin{figure}
    \centering
    \includegraphics[width=\columnwidth]{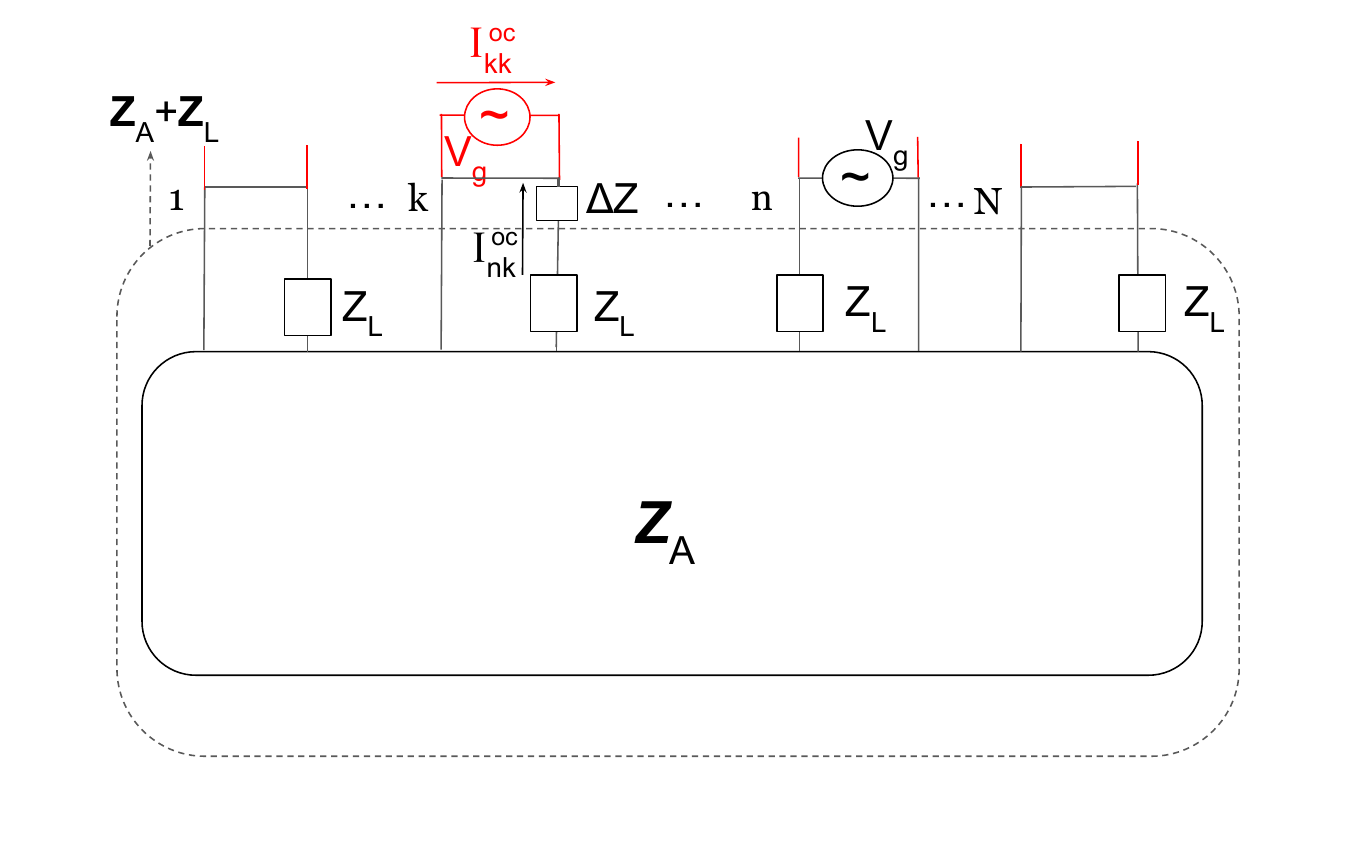}
    \caption{Microwave network schematic of a \( Z_{\rm L} \)-loaded antenna array of N elements as a N-port impedance matrix \( \mathbf{Z}_{\rm A}\), where the faulty element at position \( k \) is depicted as an extra impedance \( \Delta Z\) in series with the \( Z_{\rm L} \). The gray enclosure shows the real excitation and termination conditions, while the red one shows the self-excitation used as reference in the voltage transfer function of the second term in Eq.~(\ref{eq:EEP_equation}) (figure adapted from \cite{kyriakou2024}).}
    \label{fig:schematic}
\end{figure}

We first revisit the single-fault case of \cite{kyriakou2024} with this updated notation as well as some corrections. In an antenna array of \( N \) elements {with an impedance matrix \(\mathbf{Z}_A \in \mathbb{C}^{N\times N}\) }, such as that of Fig.~\ref{fig:schematic}, element \( k \) has a faulty termination load and instead of \( Z_{\rm L}\) it is loaded with an impedance of \( Z_{\rm F}=Z_{\rm L}+\Delta Z \). We use the Thevenin voltage source equivalent network for EEP calculation characterised by \( V_g,\ Z_g=Z_{\rm L} \). Considering the \( n\)-th element as active and following \cite{buck2023}, it is convenient to first calculate the open-circuit currents \( {\mathbf{I}_n^{\rm F,oc}}\)\(=(\mathbf{Z}_{A}+\mathbf{Z}_{\rm L}^k)^{-1}\mathbf{V}_{g,n} \)\(\in \mathbb{C}^{N\times 1}\), where \( \mathbf{V}_{g,n}=[0\hdots V_g\hdots 0]^{\rm T} \)\(\in \mathbb{C}^{N\times 1}\) with a non-zero entry only at the \( n \)-th position, while the load matrix in this case can be written as:
\( \mathbf{Z}_{\rm L}^k=Z_{\rm L}\mathbf{I}+(\Delta Z)\mathbf{u}_k\mathbf{u}_k^{\rm T}\)\(\in \mathbb{C}^{N\times N}\) where \( \mathbf{u}_k=[0\hdots 1\hdots 0]^{\rm T} \)\(\in \mathbb{C}^{N\times 1}\) a unit vector with an entry only at position \( k \). Therefore \( \mathbf{I}_n^{\rm F,oc}\)\(=(\Zbf_{\rm A}+\mathbf{Z}_{\rm L}^k)^{-1}\mathbf{V}_{g,n}=V_g\left(\left(\Zbf_{\rm A}+Z_{\rm L}\mathbf{I}+(\Delta Z)\mathbf{u}_k\mathbf{u}_k^{\rm T}\right)^{-1}\right)_{n}\). For brevity let us define the loaded admittance matrix \( \mathbf{Y}=(\Zbf_{\rm A}+Z_{\rm L}\mathbf{I})^{-1} \) of the identically loaded elements, and the loaded admittance matrix \( \mathbf{Y}^{\rm F}=(\Zbf_{\rm A}+Z_{\rm L}\mathbf{I}+(\Delta Z)\mathbf{u}_k\mathbf{u}_k^{\rm T})^{-1}\) when antenna \( k \) has a faulty termination.  By using the Sherman-Woodburry-Morrison formula {and the symmetry property \( \mathbf{Y}=\mathbf{Y}^{\rm T}\)}, the inverse matrix in the column operation can be written as:
\begin{equation}
        \mathbf{Y}^{\rm F}=\mathbf{Y}-\frac{(\Delta Z)}{1+(\Delta Z){Y}_{kk}}\mathbf{Y}\mathbf{u}_k\mathbf{u}_k^{\rm T}\mathbf{Y}=\mathbf{Y}-\frac{(\Delta Z)}{1+(\Delta Z){Y}_{kk}}\mathbf{Y}_{ k}\mathbf{Y}_{ k}^{\rm  T}
        \label{eq:inverse_rankone}
\end{equation}
\noindent
Extracting now the \( n\)-th column we have \(  \mathbf{Y}^{\rm F}_{n}\in\mathbb{C}^{N\times 1}\) {expressed as}:
\begin{equation}
        \mathbf{Y}^{\rm F}_{ n}=\mathbf{Y}_{ n}-\frac{(\Delta Z)}{1+(\Delta Z){Y}_{kk}}\left(\mathbf{Y}_{ k}\mathbf{Y}_{ k}^{\rm  T}\right)_{ n}=\mathbf{Y}_{ n}-\zeta_k{Y}_{nk}\mathbf{Y}_{ k}
        \label{eq:column_inverse_rankone}
\end{equation}
where \( \zeta_k=(\Delta Z)/(1+(\Delta Z){Y}_{kk})\). The \( n\)-th EEP of this formulation is expressed with respect to the oc-EEPs (normalized by the current source \( I_g \)) as \( E^{\mathbf{Z}_{\rm L}^k}_n=V_g/I_g(\mathbf{Y}^{\rm F}_{ n})^{\rm T}\Eoc\) \cite{warnick2021}, {where \( \Eoc\in\mathbb{C}^{N\times M}\)}. We then stack all \( N \) such row equations in a column, transposing Eq.~(\ref{eq:column_inverse_rankone}), to get \( \Ezll{k}\in \mathbb{C}^{N\times M}\):
\begin{align}
        \Ezll{k}&=\frac{V_g}{I_g}
        \begin{bmatrix}
            (\mathbf{Y}^{\rm F}_{ 1})^{\rm  T}\\
            \vdots\\
            (\mathbf{Y}^{\rm F}_{ N})^{\rm  T}
        \end{bmatrix}\Eoc
        =\frac{V_g}{I_g}\left(\begin{bmatrix}
            \mathbf{Y}_{ 1}^{\rm  T}\\
            \vdots\\
            \mathbf{Y}_{ N}^{\rm  T}
        \end{bmatrix}-\zeta_k\begin{bmatrix}
            {Y}_{1k}\mathbf{Y}_{k}^{\rm  T}\\
            \vdots\\
            {Y}_{Nk}\mathbf{Y}_{ k}^{\rm  T}
        \end{bmatrix}\right)\Eoc\nonumber\\
        &=\frac{V_g}{I_g}\left(\mathbf{Y}^{\rm T}-\zeta_k\text{diag}(\mathbf{Y}_{ k}){\underbrace{[\mathbf{Y}_{ k}\cdots \mathbf{Y}_{ k}]}_{\rm N\  times}}^{\rm T}\right)\Eoc
\end{align}
But \( V_g/I_g\mathbf{Y}^{\rm T}\Eoc=\Ezl{}\)\(\in\mathbb{C}^{N\times M}\) \cite{warnick2021} and \( V_g/I_g[\mathbf{Y}_{ k}\cdots \mathbf{Y}_{ k}]^{\rm T}\Eoc\) \(=[E_k^{\mathbf{Z}_{\rm L}} \cdots E_k^{\mathbf{Z}_{\rm L}}]^{\rm T}=E_k^{\mathbf{Z}_{\rm L}}\mathbf{1}\), where \( \mathbf{1}=[1\hdots 1]^{\rm T}\)\(\in\mathbb{C}^{N\times 1}\). Finally:
\begin{equation}
        \Ezll{k}=\Ezl{}-\zeta_k E_k^{\mathbf{Z}_{\rm L}}\text{diag}(\mathbf{Y}_{ k})\cdot\mathbf{1}=\Ezl{}-\zeta_k E_k^{\mathbf{Z}_{\rm L}}\mathbf{Y}_{ k}
        \label{eq:eep_faulty}
\end{equation}

The \( k\)-th component of these EEPs is the same as calculated in \cite{buck2023}, as, indeed, the source impedance is then `faulty' and thus different than that of the passive elements. 

Another way to express Eq.~(\ref{eq:eep_faulty}) is by introducing the open-circuit currents \( I_{nk}^{\rm oc}=Y_{nk}V_g \), which quantify the current induced in port \( n \) by excitation of a voltage source \( V_g \) on port \( k \) with all other ports open-circuited (the ports now refer to the \( \mathbf{Z}_A+{Z}_{\rm L}\mathbf{I} \) multiport network). If we consider \( \Delta Z \) as another impedance in series with \( Z_{\rm L} \) in port \( k \) (not taken into account in those \( I^{\rm oc} \) calculations), then the \( n \)-th EEP of Eq.~(\ref{eq:eep_faulty}) is equal to:
\begin{equation}
    E_n^{\mathbf{Z}_{\rm L}^k}=E_n^{\mathbf{Z}_{\rm L}}-\frac{{Y}_{kn}V_g(\Delta Z)}{V_g+{Y}_{kk}V_g(\Delta Z)}E_k^{\mathbf{Z}_{\rm L}} =E_n^{\mathbf{Z}_{\rm L}}-\frac{I_{nk}^{\rm oc}(\Delta Z)}{V_g+I_{kk}^{\rm oc}(\Delta Z)}E_k^{\mathbf{Z}_{\rm L}}
   \label{eq:EEP_equation}
\end{equation}
The factor multiplying \( E_k^{\mathbf{Z}_{\rm L}} \) is the voltage transfer function from an excitation at the now unloaded port \( n \) of the \( \mathbf{Z}_{\rm A}+{Z}_{\rm L}\mathbf{I} \) multiport network, to what is now seen as the load in port \( k \), namely \( \Delta Z \). This concept is also illustrated in Fig.~\ref{fig:schematic}, where the red drawings show the contribution of the faulty element expressed by the voltage transfer function. In the special case that \( n=k \), the resulting equation is:
\begin{equation}
    E_n^{\mathbf{Z}_{\rm L}^k}=\frac{V_g}{V_g+I_{kk}^{\rm oc}(\Delta Z)}E_k^{\mathbf{Z}_{\rm L}}
    \label{eq:circuit_divider}
\end{equation}
In this case the circuit equivalent calculates only the contribution by the open-circuit current on the excited port itself, which in circuit theory is known as a voltage divider.
Using this formula, we can straightforwardly find \( Z_{\rm F} \) from a {single measurement point (M=1) of only one of} the array \(\mathbf{Z}_{\rm L}^k\)-EEPs (under faulty termination), given that the \(\mathbf{Z}_{\rm L}\)-EEPs and the impedance matrix \( \mathbf{Z}_A \) are {all} known. Since there is only one unknown, we can take any component \( n\in{[1, N]}\) of Eq.~(\ref{eq:eep_faulty}) which leads to:
\begin{align}
    E^{\mathbf{Z}_{\rm L}^k}_n &=E^{\mathbf{Z}_{\rm L}}_n-\frac{(\Delta Z){Y}_{nk}}{1+(\Delta Z){Y}_{kk}}E_k^{\mathbf{Z}_{\rm L}}\Rightarrow \label{eq:eep_faulty_n}\\
    Z_{\rm F} &=Z_{\rm L}+\frac{E^{\mathbf{Z}_{\rm L}}_n-E^{\mathbf{Z}_{\rm L}^k}_n}{{Y}_{nk}E^{\mathbf{Z}_{\rm L}}_k-{Y}_{kk}(E^{\mathbf{Z}_{\rm L}}_n-E^{\mathbf{Z}_{\rm L}^k}_n)}
    \label{eq:Z_F}
\end{align}
{Even though in sampling the electric far-field quantities, only one measurement point is needed for this equation to work}, multiple points might nonetheless offer a more robust estimation when measurement uncertainties impact the calculations, by solving a least squares problem as will be shown in the next section.

It is important to also note that the equations presented are agnostic to the form of \( \mathbf{Y}\) as the series connection of an antenna impedance matrix with a scaled unity matrix; the perturbation approach also works when applied to a matrix of any other form. We can therefore generalise the single-fault case to achieve impedance fault extraction for any number of faulty loads, by exploiting the same formulation when the matrix \(\mathbf{Y} \) is continuously updated from its previous form.

\IEEEpubidadjcol
\section{EEPs with N faulty terminations and recursive inversion algorithm \label{sec:new_algorithm}}
\noindent
Let us now suppose that there are \( N \) faults with respect to the nominal termination, \( (\Delta Z)_k \), \( k\in{[1, N]} \), the same as the number of antennas. A similar calculation of the EEPs for this case follows a recursive procedure both for Eq.~(\ref{eq:inverse_rankone}) and Eq.~(\ref{eq:EEP_equation}). We first introduce the notation: 
\begin{equation} \mathbf{\Gamma}^k=\zeta_k\mathbf{Y}^{{k-1}}_{ k}
\label{eq:Lambda}
\end{equation}
which is a dimensionless quantity and akin to a reflection coefficient, as will also be seen from the final equation. The superscript `F' is now dropped from the admittance matrix and replaced by the iteration number. We then follow Eq.~(\ref{eq:inverse_rankone}) recursively, defining:
\begin{equation}
    \begin{cases}
    \mathbf{Y}^0&=\mathbf{Y}\\
    \mathbf{Y}^{1}&=\mathbf{Y}^{0}-\mathbf{\Gamma}^{1}(\mathbf{Y}^0_{ 1})^{\rm T}\\
    &\hdots\\
    \mathbf{Y}^N&=\mathbf{Y}^{N-1}-\mathbf{\Gamma}^{N}(\mathbf{Y}^{N-1}_{ N})^{\rm T}
    \end{cases}
\end{equation} 
The general recursive equation is, therefore,:
\begin{equation}
    \mathbf{Y}^{k}=\mathbf{Y}^{k-1}-\mathbf{\Gamma}^{k}(\mathbf{Y}^{k-1}_{k})^{\rm T}
    \label{eq:Y_recursive}
\end{equation}

\begin{figure}
    \centering
    \includegraphics[width=\columnwidth]{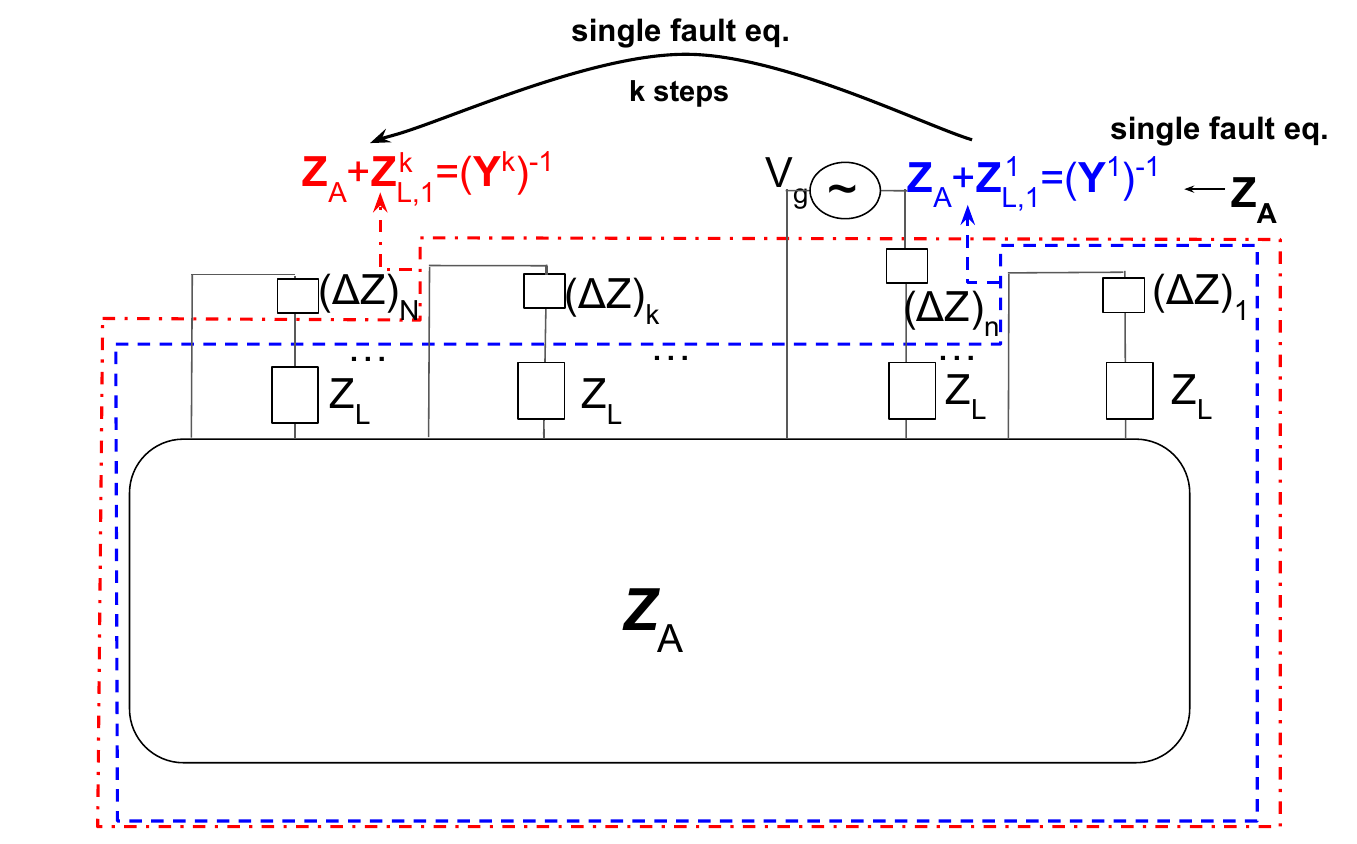}
    \caption{Schematic representation of the recursion with respect to its progression using impedance matrices. The encircled blocks are defined as impedances, while also expressed as the inverse of the respective admittance matrix presented in the text. Element \( n \) radiates, while element \( k \) is a sequential element of the recursion.}
    \label{fig:eep_recursion}
\end{figure}

We now proceed to calculate recursively the EEPs. First, we introduce the notation: \( \mathbf{Z}_{{\rm L},1}^{k}\) to signify a loading matrix where elements \( 1 \) through \(k\) are faulty. We simplify our previous EEP notation defining: \( \Ezll{}=\mathbf{E}^{0},\ \mathbf{E}^{\mathbf{Z}_{{\rm L},1}^1}=\mathbf{E}^{1},\hdots,\ \mathbf{E}^{\mathbf{Z}_{{\rm L},1}^N}=\mathbf{E}^{N} \). This numbering agrees with the previous admittance matrix numbering, and therefore the \( \mathbf{\Gamma} \) numbering as well. Fig.~\ref{fig:eep_recursion} graphically shows how the recursive computation of the new loaded impedance matrices progresses on a schematic of the array loaded with arbitrary loads \( Z_{\rm L}+(\Delta Z)_k,\ k\in{[1, N]}\). Equipped with these definitions, we can write down the general, recursive equation based on Eq.~(\ref{eq:eep_faulty}) as:
\begin{equation}
    \mathbf{E}^{k}=\mathbf{E}^{k-1}-\mathbf{\Gamma}^{k}E^{{k-1}}_k
\end{equation}

We subsequently consider successively the \( n\)-th component of \( \mathbf{E}^{n} \), \( n\in{[1, N]} \), and write down Eq.~(\ref{eq:EEP_equation}): 
\begin{equation}
    \begin{cases}
    E_1^1&=E_1^0-\Gamma^1_{1} E_1^{0} \\
    E_2^2&=E_2^1-\Gamma^2_{2} E_2^{1} \\ &=E_2^0-\Gamma^1_{2} E_1^{0}-\Gamma^2_{2} E_2^0+\Gamma^2_{2}\Gamma^1_{2} E_1^0\\
    &=E_2^0-\Gamma^2_{2} E_2^{0}-\Gamma^1_{2}(1-\Gamma^2_{2})E_1^0 \\ &\hdots \\ E_N^N&=E_N^{N-1}-\Gamma_N^NE_{N}^{N-1} \\
    &=E_N^{0}-\Gamma^N_{N}E_N^0-\hdots-\Gamma^1_{N}\left(\prod\limits_{l=2}^{N}(1-\Gamma^l_{N})\right)E_1^0\\
   
    \end{cases}\nonumber
\end{equation}

The last, general equation is proven by induction in Appendix~\ref{sec:appendix} and can be written in a more compact form as:

\begin{equation}
 E_N^N=E_N^0-\sum\limits_{k=1}^N\Gamma^k_{N}\left(\prod\limits_{l=k+1}^{N}(1-\Gamma^l_{N})\right)E_k^0 \label{eq:recursive_EEPs}
\end{equation}

This `forward' equation shows that the \( \Gamma_N^k \) and therefore \( \zeta_k \), \( (\Delta Z)_k \) for all \( k\in \) \([1, N]\) are used for the calculation of faulty EEP \( E_N^N \). This equation does not provide any advantage with respect to forwardly calculating the EEPs given certain impedance faults, but sets the stage for an algorithm which can perform the inverse calculation, namely calculating the faults given certain sets of EEPs.

We furthermore note that the ordering of the antennas in the above derivation does not matter, and the same expressions apply with different orderings. This is a concept that is best interpreted through signal flow graphs, but proving such an analogy is out of the scope of this paper. However, this observation means the `last', \( N\)-th antenna is free to choose, so we will generalise our expressions by replacing \( N\) with \( n\), keeping in mind that every time we choose a different reference antenna, some permutation of rows/columns of the respective matrices has to be performed, in order to place it in the \( N\)-th position {(see first step of Alg.~\ref{alg:termination_fault})}. 

To facilitate devising a strategy to calculate \( (\Delta Z)_k \), we notice that Eq.~(\ref{eq:recursive_EEPs}) is written simply as:

\begin{equation}
    E_n^0-E_n^N=\sum_{k=1}^NT_{k} E_k^0
    \label{eq:linear_combination}
\end{equation}
where \( T_{k}=\Gamma^k_{n}\left(\prod_{l=k+1}^{N}(1-\Gamma^l_{n})\right) \) {for a certain choice of \(n\) which we here omit in the \( T\) notation to avoid indexing confusion}. If we know the EEP values at an initial (nominal) and a final (faulty) loading condition at {\(M\)} sampled far-field points, the coefficients \( T_{k} \), which are unknown, can be solved by setting up a system of linear equations when we have at least \( N \) such far-field points and solving it using the pseudoinverse matrix (least-squares projection). Then the \( \Gamma^k_{n} \) coefficients can be retrieved in reverse starting from \( \Gamma_n^N=T_{N} \) and following backward substitution for \( k=N-1,\hdots,\ 1 \) as:

\begin{equation}
   \Gamma^k_{n}=\frac{T_{k}}{\prod_{l=k+1}^{N}(1-\Gamma^l_{n})}
\end{equation}

Finally, we can extract the faults by the definition of \( \zeta_k \) and taking the \( n \)-th component of Eq.~(\ref{eq:Lambda}) as:

\begin{equation}
    (\Delta Z)_k=\Gamma^k_{n}/({Y}^{k}_{kn}-\Gamma^k_{n}{Y}^k_{kk})
\end{equation}

This is done in incremental numbering, since each \( (\Delta Z)_k \) is needed to calculate the new \( \mathbf{Y}^{k+1} \) via Eq.~(\ref{eq:Y_recursive}), including the knowledge of all pre-calculated \( \Gamma_n^k,\ k\in{[1, N]}\). We only use the \( n \)-th \( \mathbf{\Gamma}^k \) component in {that last algorithmic step}, but we then implicitly define all the \( n-1 \) remaining components to continue the recursion. The algorithm is presented in the form of pseudocode in Alg.~\ref{alg:termination_fault}. The vector (overbar) quantities always refer to a stacking of many (in general \( M\geq N\)) far-field points along the row dimension, while the column dimension in the matrix quantities has been used to stack the different antenna patterns.

\begin{algorithm}

\caption{Inversion algorithm of Eq.~(\ref{eq:recursive_EEPs}) }\label{alg:termination_fault}
{\texttt{Inputs:} $\bar{E}_k^0 \in\mathbb{C}^{1\times M}\ \forall k\in[1, N]$, $\bar{E}_n^N$ \texttt{for selected} $n$}\\
{\texttt{Outputs:} $(\Delta Z)_k\ \forall k\in[1, N]$}\\ 
{$ \mathbf{s} \gets [\texttt{randomPermutation}(\{1,\hdots,n-1,n+1,\hdots,N\})\ n]$} \\
{$\mathbf{Z}_A\gets[Z_{A,{s_{1}s_{1}}}\cdots Z_{A,s_{1}s_{N}}; \cdots; Z_{A,s_{N}s_{1}}\cdots Z_{A,s_{N}s_{N}}] $} \\
$\bar{\mathbf{E}}^0\gets[\bar{E}^0_{s_{
1}};\cdots;\bar{E}^0_{s_{N}}] $ \\
$\mathbf{T} \gets \left((\bar{\mathbf{E}}^{0})^{\rm {H}}\bar{\mathbf{E}}^{0}\right)^{-1}(\bar{\mathbf{E}}^{0})^{\rm {H}}(\bar{E}_n^0-\bar{E}_n^N)$ \\
$\Gamma^N_{n}\gets T_{N}$\\
\textbf{for }{$k=N-1,1,1$}\\ 
    \hspace*{4mm}
    $\Gamma^k_{n}=\frac{T_{k}}{\prod\limits_{l=k+1}^{N}(1-\Gamma^l_{n})} \nonumber$
\\ 
\textbf{end}\\
\( \mathbf{Y}^{0}=\left(\mathbf{Z}_{\rm A}+Z_{\rm L}\mathbf{I}\right)^{-1}\)\\
\textbf{for }{$k=1,1,N-1$}\\ 
    \hspace*{4mm} $(\Delta Z)_k \gets \Gamma^k_{n}/({Y}^{{k-1}}_{kn}-\Gamma^k_{n}{Y}^{{k-1}}_{kk})$ \\ 
    \hspace*{9.2mm} $\mathbf{\Gamma}^k \gets (\Delta Z)_{k}/\left(1+(\Delta Z)_{k}{Y}^{{k-1}}_{kk}\right)\mathbf{Y}^{{k-1}}_{k}$ \\ 
    \hspace*{5.4mm} $\mathbf{Y}^{k} \gets \mathbf{Y}^{k-1}-\mathbf{\Gamma}^{k}(\mathbf{Y}^{{k-1}}_{k})^{\rm  T}$\\ 
\textbf{end}\\
$(\Delta Z)_N \gets \Gamma^N_{n}/({Y}^{{N-1}}_{Nn}-\Gamma^N_{n}{Y}^{{N-1}}_{NN})$ \\
$ $
\end{algorithm}

It is worth mentioning that since ultimately every component \( E_n^N \) of the final EEPs can be expressed as a linear combination of \( E_1^0,\hdots,\ E_N^0\), after retrieving the \( T_{k,n}\) coefficients, the problem boils down to determining \( (\Delta Z)_1,\hdots,(\Delta Z)_N \) now as a non-linear function of these entries \( T_{k} \) and the {entries of} admittance matrix \( \mathbf{Y}\). This is a computationally hard problem due to the inverse matrix operation involved in the matrix transformation equations of \cite{kovaleva2020}. Symbolic math in numerical codes which can perform such algebra to define the exact non-linear function can easily  result in the phenomenon known as `expression swell' for large enough arrays demanding exponentially large calculation costs. Therefore, the recursive approach presented here is well-suited to solving this problem with computational efficiency.

Finally, the algorithm is best implemented when all elements are considered as potentially having a termination fault, and those who do not present such deviation from the nominal value will result in \( \Delta Z =0\ \)\( \Omega\) up to a certain tolerance.

\section{Numerical Validation}
\subsection{Simulated patterns without noise \label{sec:simulation_nonoise}}
We will first test our algorithm with a simulated array to establish possible numerical errors, which are known to accumulate in recursive computations. We also remind the reader that any single EEP of those under faulty termination can be used as the right hand side of Eq.~(\ref{eq:linear_combination}), so this parameter will also be assumed free, in order to examine the effect of such a `reference' EEP and whether the particular geometric placement of its corresponding array element has any effect. Let us denote this `reference' antenna number \( n \) {(measurement index)}, while the number of the antenna for which the new termination is estimated is denoted \( k\) {(fault index)}. The true terminations are denoted \( Z_{{\rm F},k}^{\rm true}\), while the estimated ones are \( Z_{{\rm F},k}^{{\rm alg},n}\).

As in \cite{kyriakou2024}, we will use as an example a 16-element tile of the Murchison Widefield Array (MWA) \cite{wayth2018}. This radio astronomical array is located at Inyarrimanha Ilgari Bundaya, a radio quiet zone in Western Australia, and conducts low-frequency science in the 70 to 300~MHz band. The electromagnetic characterisation for such a tile, which is a \( 4\times 4\) subarray within the instrument, has extensively been treated in \cite{sutinjo2015}. We will consider the MWA tile EEPs at 128~MHz, {simulated with FEKO}, with a nominal termination of \( Z_{\rm L}=50~\Omega\), while faults will be introduced at 4 positions of the array, such that: \( Z_{\rm F,1}^{\rm true}=28.87+j15.98~\Omega,\ Z_{\rm F,6}^{\rm true}=19.9+j11.01~\Omega,\ Z_{\rm F,11}^{\rm true}=21.6+j11.96~\Omega\) and \( Z_{\rm F,16}^{\rm true}=13+j7.2~\Omega\). These correspond to real measurements of faulty LNA input impedances mounted on the MWA tile elements \cite{kovaleva2020}.

\begin{figure}
    \centering
    \includegraphics[width=\columnwidth]{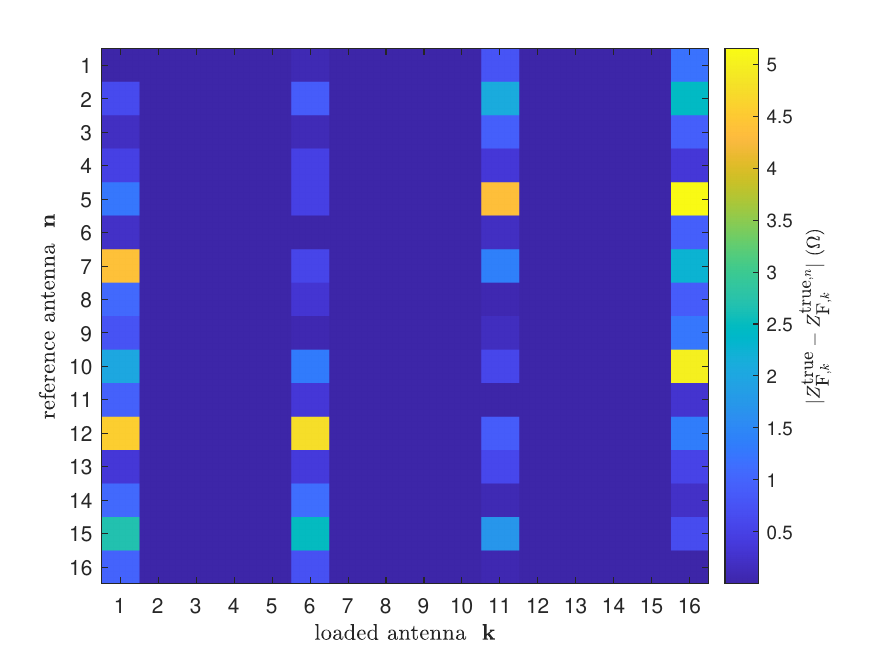}
    \caption{Discrete color grid of the absolute error value using the far-field sampling \( \Omega_\theta\times\Omega_\phi \), of the estimations of the faulty termination impedance of antenna \( n \) using the reference EEP of antenna \( k \). }
    \label{fig:mwa_fault_over_pivot}
\end{figure}

Fig.~\ref{fig:mwa_fault_over_pivot} shows the absolute error between true and algorithmically estimated values of the faulty termination impedance of antenna \( k \) using the reference EEP of antenna \( n \) and a sampling of the 3D far-field described by the sets \( \Omega_\theta=\{n_\theta\cdot 5^\circ,\ n_\theta=0,\cdots,9\} \textrm{ and } \Omega_\phi=\{n_\phi\cdot 45^\circ,\ n_\phi=0,\cdots,7\}\). As can be seen from this plot, higher errors are observed when using the 5-th, 7-th or 12-th antenna as reference, while all other choices generally lead to errors of less than \( 4\ \Omega\). The best reference antenna is the 6-th one, with an RMS error over all faults equal to 0.21 \( \Omega \) (or {0.5\%} normalized). It is also observed that the algorithm always predicts where a fault does not exist: if we exclude columns 1, 6, 11, 16 from Fig.~\ref{fig:mwa_fault_over_pivot} (where the loaded antennas do have a fault), then the maximum error is \( 1.5\times 10^{-7}\ \Omega\). {It has also been observed that more faulty impedances increase the normalized estimation error, but increasing the termination faults corresponds to increasingly non-realistic cases.}

As far as the computation time is concerned, as can be seen from Alg~\ref{alg:termination_fault}, the complexity is \( \mathcal{O}(N^3) \), defined both by the \( \mathbf{Y}^0 \) computation as well as the \(N \) iterations of the \( \mathbf{\Gamma}^k(\mathbf{Y}^{k-1}_k)^{\rm T}\) computation (we consider the standard and not accelerated matrix operation complexities). Using Matlab on a Lenovo laptop with an Intel Core 7, 10-core processor at 1.8~GHz and 32~GB of RAM, the time elapsed for one full computation is 0.0062 seconds. Attempts at solving the problem using optimisation methods on the non-linear expression of \( T_{k} \)'s derived through symbolic math and substituted to Eq.~(\ref{eq:linear_combination}) did not give convergent results, while at the same time requiring a computation time between 15-30 seconds.

\subsection{{Simulated patterns: additive, multiplicative noise and other measurement systematics}  \label{sec:simulation_noise}}
Since the method presented is of potential use with measurement data, the case of non-ideal EEPs will be examined. The scattering matrix will in this section also be considered fully known; this ideal condition should in fact also be further examined according to errors that can occur in indirect determination of this matrix, as has been studied in \cite{kyriakou2024b}. In order to model a potential measurement of an EEP, we introduce a complex Gaussian additive noise term, which is drawn from the Gaussian distribution parametrised by the Signal-to-Noise Ratio (SNR). We furthermore implement 1000 realizations of that term per each SNR point examined and average the resulting squared errors computed, so as to ensure statistically unbiased results. Thus, the noise term is described {for any \( \omega=(\theta,\phi)\)} as:
\begin{equation}
    w_k^m{(\omega)}\sim\mathcal{N}_r(0,\sigma_k^m{(\omega)})+j\mathcal{N}_i(0,\sigma_k^m{(\omega)}),\ \sigma_k^m{(\omega)}=\sqrt{\frac{{|E_k^m(\omega)|^2}}{\rm SNR}}
\end{equation}
where the \( k\)-th EEP {\( E_k^m,\ m\in\{0,n\}\) is used (`0' nominal and `\(n\)' for faulty)} and the \( r,\ i \) indices of the normal distribution random variables signify different realisations of the real and imaginary part (seeds in statistical terminology). {The next step is to introduce multiplicative noise, which models the fading nature of the line-of-sight channel when a transmitter probe and the antenna-under-test as receiver interact in a measurement scenario. Such behavior is modelled by a Rician distribution of the multiplicative factor \( g_k\sim\mathcal{R}(\nu_k,\sigma_{R,k})\) that affects the amplitude of the received embedded element pattern \( E_k^m\). The phase is here assumed to be known and better controlled, as various techniques ensure its correct retrieval with respect to the reference antenna \cite{ciorba2022}. As has also been practiced in \cite{kyriakou2024b}, we can use a normalised channel amplitude factor so as to have \( \mu_{R,k}=1 \). We examine this noise by setting the fading level of the channel, namely \( {\rm K}=\nu_k^2/(2\sigma_{R,k}^2)\). From these two relationships and the Rician distribution formula for the \( g_k \) mean value we can derive \( \nu_k,\ \sigma_{R,k} \).  If \( \hat{E}_k^m\) is the measured EEP, our model now reads:
\begin{equation}
    \hat{E}_k^m(\omega)=g_kE_k^m(\omega)+w_k^m(\omega)
\end{equation}} 
{For the \( {\rm K}\rightarrow\infty \) case}, the EEP chosen from the array at faulty conditions, \( n\), will be examined both by averaging over all possible elements and by choosing the one that minimises the current error, that is, the RMS of errors of the estimated termination impedances at a certain SNR. This antenna will be selected as the one which most frequently minimises such an error across all 1000 realisations of the computation. In these realisations, we also randomly re-arrange all other elements in positions \( 1 \) through \( N-1\) (while placing the \( n \)-th at position \( N \)), in order to get an impartial result independent of the MWA array element numbering. {The following results differ also in the \( \phi \) sampling, which was increased to 16 azimuthal cuts that ensure faster convergence.}

\begin{figure}
    \centering
    \includegraphics[width=\columnwidth]{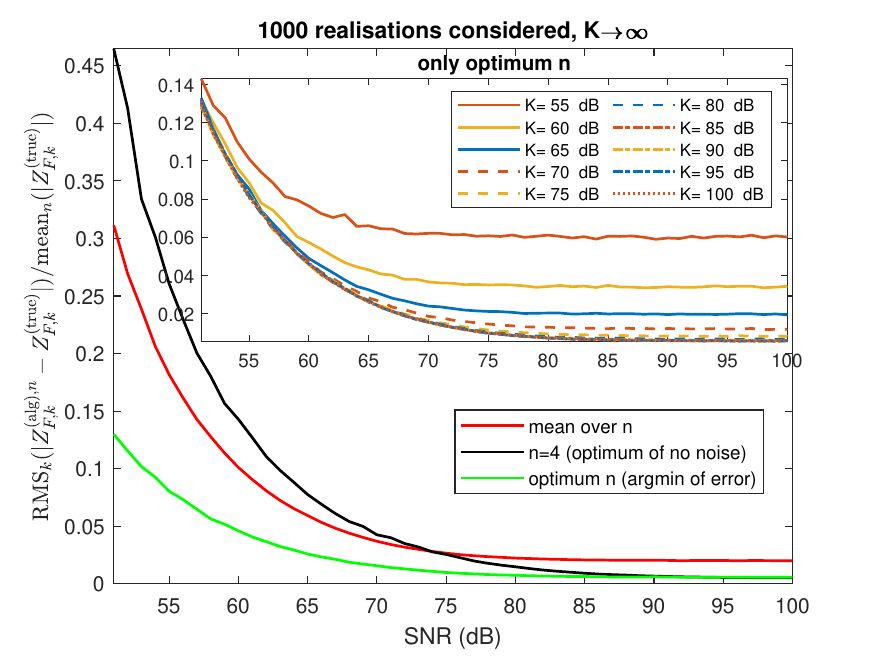}
    \caption{RMS error versus SNR of true with respect to algorithmically estimated termination impedance, normalised over the mean of all impedances across the SNR range, for a 16-element MWA tile over 1000 realisations of additive Gaussian noise {and multiplicative Rician fading}. Results {when \( {\rm K}\rightarrow\infty\)} are shown averaged over computations for {all} reference element{s} \(n\), for the case of \( n=4\) and for the optimum EEP \(n\) that most frequently minimises the RMS across all 1000 realisations. An inset also shows the SNR trend of the {`optimum n' case of all other fading levels, where each \( \rm K \) curve is reported in the inset legend.} }
    \label{fig:mwa_RMSE_vs_SNR}
\end{figure}

Fig.~\ref{fig:mwa_RMSE_vs_SNR} presents the RMS of the estimations of the faulty termination impedance over all antennas as a function of the SNR (in dB) {from 50 to 100 for \({\rm K}\rightarrow\infty\) (no fading), while a sampling of \( {\rm K}\) (in dB) at sparser values (from 55 to 100 with a step of 5) is shown for the `optimum \( n \)' case in an inset}. The result is normalised over the mean value of the \( |Z_{{\rm F},k}^{\rm true}|,\ k\in \) \([1, N]\). The case of \( n=4 \) is also plotted {when  \( {\rm K}\rightarrow\infty \) } for comparison, as it has been found to be the optimum \( n \) in the noiseless EEP case.

It can be seen that, when computing the normalized RMSE in an average sense over all possible reference elements, the error {converges slower to acceptable values, such as the crossing of 5\% at SNR\(>65\)~dB}. When a fixed element that minimizes the error is chosen to perform the computation, the error {converges much faster, roughly at SNR\(>58\)~dB for the 5\% threshold. This element is always the 6-th, as has been verified by keeping the `argmin' indices during the simulation, while the 11-th also presents similar performance}. The \( n=4\) case {presents a behavior more similar to the `mean over \( n\)' curve, highlighting the necessity of this noise study; indeed, in the noiseless case, that element outperformed the ones highlighted here by a small margin}. According to the original numbering of an MWA tile, {the 6-th and 11-th elements lie in the inner part of the rectangular array configuration and as has been assumed in this numerical experiment, are within the elements that present an impedance fault.} {It can also be seen that, as with the case of no fading (\( {\rm K}\rightarrow \infty\)), for SNR\(>60 \)~dB and a finite \( {\rm K}\geq 60\) dB the error converges to ever lower values by increasing \( {\rm K} \). In contrast to \cite{kyriakou2024b}, there is now a clear trend of the RMSE asymptot value diminishing by {further} increasing K. It is therefore equally important to keep both SNR and K higher than 60 dB to obtain results within 5\% accuracy.}

\begin{figure}
    %\centering
    \hspace{-1.5em}
    \subfloat[]{\includegraphics[width=0.55\columnwidth]{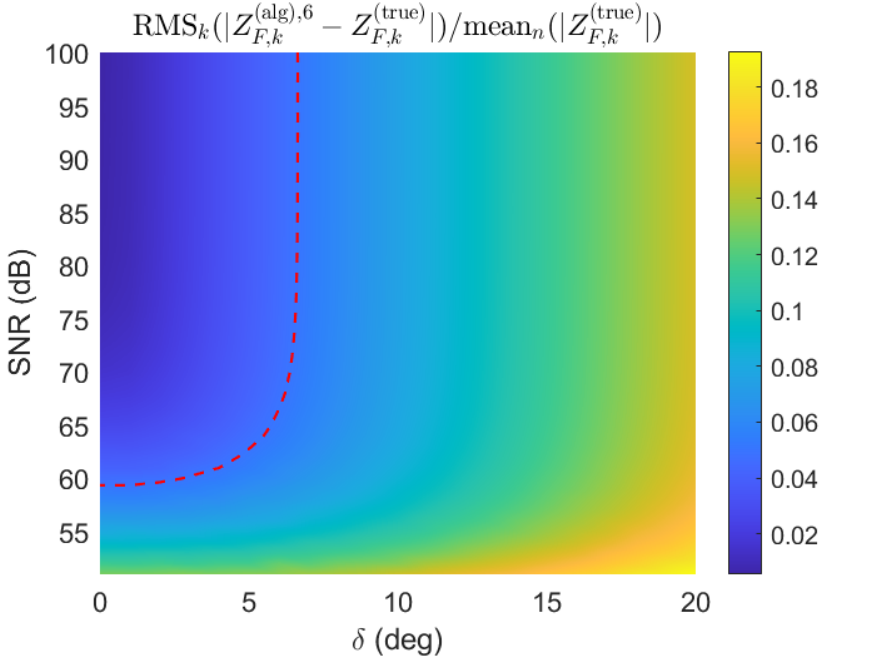}\label{fig:mwa_RMSE_vs_SNR+phase}}
    \quad \hspace{-2.2em}
    \subfloat[]{\includegraphics[width=0.56\columnwidth]{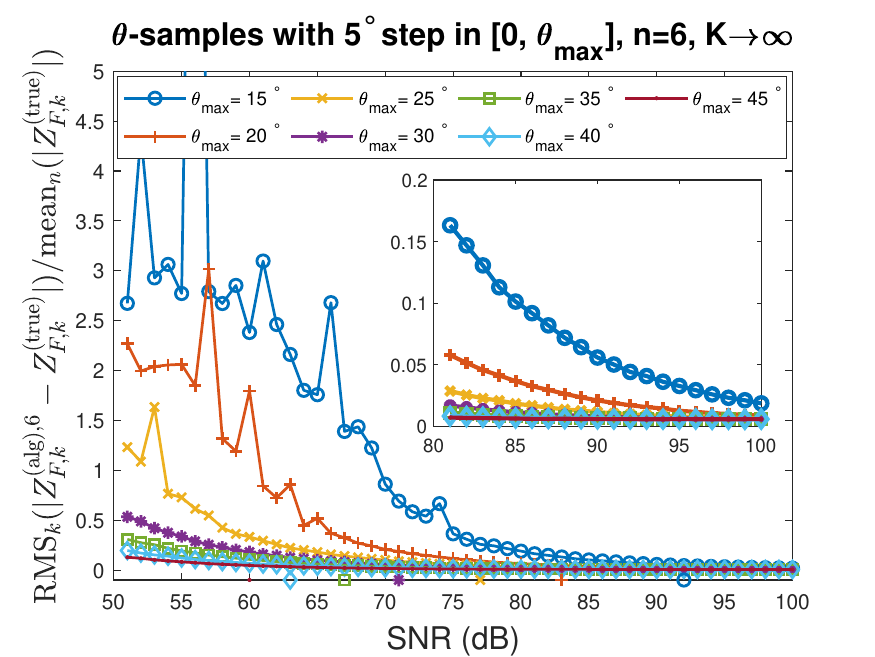}\label{fig:mwa_RMSE_vs_SNR+sampling}}
    \caption{{Normalized RMSE for \(\rm K\rightarrow\infty\), \( n=6\): (a) versus SNR and an induced phase difference \(\delta \) between \(\bar{E}_n^0\), \(\bar{E}_n^N \), (b) versus SNR across different numbers of samples \( \theta\in [0, \theta_{\rm max}]\) by increasing \( \theta_{\rm max} \) while keeping the \( \theta\) sampling at \( 5^\circ\) and the \( \phi \) one such that \(\Omega_\phi=\{n_\phi\cdot22.5^\circ,\ n_\phi=0,\hdots,15\}\).}}
\end{figure}

Finally, a brief parametric study is performed with respect to systematics known to impact far-field pattern measurements. We focus our attention on the far-field pattern sampling, which can be limited in real scenarios, as well as the phase drift between acquisitions of far-field patterns across different measurement campaigns. We restrict this study to the case of \( {\rm K}\rightarrow\infty\) and the \( n=6\), as the identified element that minimizes the RMSE. In Fig.~\ref{fig:mwa_RMSE_vs_SNR+phase}, the normalized RMSE across SNR is calculated simulating possible phase differences \( \delta \) between the \( \bar{E}_n^0\), \(\bar{E}_n^N\) quantities which would be acquired by different measurement campaigns, ranging from \( 1^\circ \) to \( 20^\circ \) with a \( 1^\circ \) step, having the least-squares step of Alg.~\ref{alg:termination_fault} being solved with a right-hand-side \( (\bar{E}_n^0-e^{j\delta}\bar{E}_n^N)\). In Fig.~\ref{fig:mwa_RMSE_vs_SNR+sampling}, the normalized RMSE across SNR is now examined with respect to the sampling within a limited zenith angle range by keeping 16 azimuthal planes and varying the number of samples of the \( \Omega_{\theta}\) space using the same \( 5^\circ \) step and increasing the maximum angle \( \theta_{\rm max}\) from \( 15^\circ \) up to \( 45^\circ \). We note a certain degree of robustness of the algorithm with respect to these two systematics: up to \( \delta=7^\circ\), the RMSE is still below the 5\% threshold (red dotted line) at around 60-65~dB, whereas for larger \( \delta \) this threshold becomes asymptotically out of reach for however large SNR values; yet, the RMSE is still mostly kept within 10-15\%. On the other hand, the \( \theta \) sampling that ensures an already fast rate of convergence is that of \( \theta_{\rm max}=30^\circ\), a zenith angle that in practical measurements would not suffer from other non-line-of-sight effects. Each curve marker on the \(x\)-axis indicates the 5\% threshold point, seen at 92~dB down to 60~dB, as \( \theta_{\rm max}\) increases from \( 30^\circ\) up to \( 45^\circ \). 
\vspace{-2em}
\section{Conclusions \label{sec:conclusions}}
An algorithm that computes impedance faults of the front-end termination of \( N \)-element array antennas has been described, using a full set of EEPs at the nominal termination impedance, and one EEP at non-uniform (faulty) termination. Using simulated EEPs from a 16-element tile of a radio-astronomical array, the algorithm correctly finds the impedance fault at an accuracy of around {0.5\%}; that error is higher when noise is added in our model to estimate how a measured EEP would affect the calculation and the algorithm then {crosses the 5\% error only} for high signal-to-noise ratio and fading channel levels {(both \(\sim 60-65\)~dB)}. In our example, {the inner faulty elements} of the rectangularly arranged array are the ones that achieve the minimum error when their EEP is chosen as the right-hand-side of the least-squares problem with which the algorithm starts; more configurations would need to be tested to ascertain whether such elements are always the optimum choice for our proposed method. Our results show that the algorithm can reliably be applied in radio astronomical antenna arrays for fault detection of individual elements' front-end impedance using the minimum number of EEP measurements, {handling at the same time various systematic limitations}.
\vspace{-1em}
[Proof of Eq.~(\ref{eq:recursive_EEPs})]
{\appendix
\label{sec:appendix}
We prove here the general form of Eq.~(\ref{eq:recursive_EEPs}) for a \( N \)-element array using induction, where \( m \) is the integer number of the recursion step. For \( m=1 \), the product is by definition equal to 1 (since the upper limit is lesser than the lower limit) and the equation reduces to:
\begin{equation}
    E_n^1=E_n^0-\sum\limits_{k=1}^1\Gamma^k_{n}\left(\prod\limits_{l=k+1}^{1}(1-\Gamma^l_{n})\right)E_k^0=E_n^0-\Gamma^1_{n}E_1^0
\end{equation}
This is the equation of the single case, proven in Sec.~\ref{sec:eep_faulty}. We now suppose that Eq.~(\ref{eq:recursive_EEPs}) holds when the integer step \( 1\leq m\leq N-1\) is considered for all \( n\in \) \(  [1, N]\), and examine its form for \( m+1\):
\begin{align}
    E_{n}^{m+1}&=E_{n}^{0}-\sum\limits_{k=1}^{m+1}\Gamma^k_{n}\left(\prod\limits_{l=k+1}^{m+1}(1-\Gamma^l_{n})\right)E_k^0 \nonumber \\
    &=E_n^0-\Gamma_n^{m+1}E_{m+1}^0-\sum\limits_{k=1}^{m}\Gamma^k_{n}\left(\prod\limits_{l=k+1}^{m+1}(1-\Gamma^l_{n})\right)E_k^0 \nonumber
\end{align}
Since for \(k\leq m \Rightarrow k+1\leq m+1\), all terms of the sum have the factor \( 1-\Gamma_n^{m+1} \) as common, so the above equation takes the form:
\begin{align}
    E_n^{m+1}&=E_n^0-\Gamma_n^{m+1}E_{m+1}^0 - \sum\limits_{k=1}^{m}\Gamma^k_{n}\left(\prod\limits_{l=k+1}^{m}(1-\Gamma^l_{n})\right)E_k^0\nonumber\\ 
    &+\Gamma_n^{m+1}\sum\limits_{k=1}^{m}\Gamma^k_{n}\left(\prod\limits_{l=k+1}^{m}(1-\Gamma^l_{n})\right)E_k^0
\end{align}
By virtue of the assumption that the equation hold for integer \( m \), the first plus third terms can be substituted by \( E_n^m\) so that:
\begin{equation}
    E_n^{m+1}=E_n^m-\Gamma_n^{m+1}\left(E_{m+1}^0- \sum\limits_{k=1}^{m}\Gamma^k_{n}\left(\prod\limits_{l=k+1}^{m}(1-\Gamma^l_{n})\right)E_k^0\right) %\nonumber
\end{equation}
We now set \( n=m+1 \) and note that the term in parenthesis is again by the same assumption equal to \( E_{m+1}^m \), such that we obtain:
\begin{equation}
    E_{m+1}^{m+1}=E_{m+1}^m-\Gamma_{m+1}^{m+1}E_{m+1}^m
\end{equation}
But this equation is the general, rank-one update on the (\( m+1\))-th EEP, which has been proven in Sec.~\ref{sec:eep_faulty}. Eq.~(\ref{eq:recursive_EEPs}) in the main text is the application of this equation when \( m+1=N \) but it has been claimed to be valid for all \( n \). As has been emphasized in the text, {we can now place another element as the \(N\)-th one and,} since the induction assumption is valid for any \( n\in \) \([1, N]\), the above procedure can be repeated without further assumptions, where the \( \Gamma^k_{m+1},\ k\in \) \( [1, N]\) are now different factors. Hence the proof is complete.
}
\vspace{-1em}
\section*{Acknowledgments}
I would like to thank Karl F. Warnick and Tobia Carozzi for having read earlier versions of the manuscript and provided useful comments, {as well as the anonymous reviewers for their constructive reviews}. I also thank Maria Kovaleva for having provided the MWA tile simulated EEPs. I acknowledge the Wajarri-Yamatji people as the traditional owners of the observatory site where Murchison Widefield Array operates; and all indigenous peoples' right to their land.

\vfill

\end{document}